\documentclass[12pt]{article}

\usepackage[a4paper,left=30mm,right=20mm,top=20mm,bottom=20mm]{geometry}
\usepackage{graphics}
\usepackage{amsmath}
\usepackage{amssymb}
\usepackage{epsfig}
\usepackage{cite}
\usepackage{xcolor}
\usepackage[unicode]{hyperref}
\newcommand{\bea}{\begin{eqnarray}}
\newcommand{\eea}{\end{eqnarray}}
\newcommand{\be}{\begin{equation}}
\newcommand{\ee}{\end{equation}}

\title{A simple model for PDFs and nPDFs
}

\author{A.V.~Kotikov$^{1}$, A.V.~Lipatov$^{2}$}

\begin{document}

\maketitle

\begin{center}
{\it $^{1}$Joint Institute for Nuclear Research, 141980 Dubna, Moscow region, Russia}\\
{\it $^{2}$Skobeltsyn Institute of Nuclear Physics, Lomonosov Moscow State University, 119991 Moscow, Russia}

\end{center}

\vspace{0.5cm}

\begin{center}

{\bf Abstract }

\end{center}

\indent

We present the main results of our recent papers \cite{KL2025,KL2025res}. In
\cite{KL2025}, we derived an analytical solution of the QCD evolution equations for
parton distribution functions.
The valence and non-singlet quark components
satisfy the Gross-Llewellyn-Smith and Gottfried sum rules, respectively, while
momentum conservation is maintained for the singlet quark and gluon densities.
Several phenomenological parameters were extracted from a combined fit to precision
data on the proton structure function $F_2(x,Q^2)$ collected by the BCDMS, H1, and
ZEUS Collaborations, comprising a total of 933 points from 5 datasets. In
\cite{KL2025res}, we proposed a model for nuclear medium modifications of parton
densities. The approach is based on a global analysis of available deep inelastic
scattering data for different nuclear targets within the rescaling model,
incorporating Fermi motion effects. By fitting the rescaling parameters to
experimental data on the ratio $F_2^A(x,Q^2)/F_2^{D}(x,Q^2)$ for several nuclear
targets $A$, we obtained predictions for nuclear parton distribution functions.

\vspace{1.0cm}


\vspace{1.0cm}

\newpage

\section{Introduction} \indent

The study of deep inelastic scattering (DIS) of leptons on nuclei reveals
significant effects of nucleon interactions within the nucleus, challenging the
naive picture of the nucleus as a system of quasi-free nucleons (see the recent
review \cite{MKla24}). Nuclear medium effects on parton distribution functions
(PDFs) attract considerable interest from both experimental and theoretical
perspectives. A detailed understanding of nuclear modifications of parton densities
(nPDFs) is a fundamental problem in nuclear and high-energy physics, crucial for
theoretical descriptions of $pA$ and $AA$ collisions at modern (LHC, RHIC) and
future colliders (FCC-he, EiC, EicC, CEPC, NICA). The nuclear modification factor is
typically defined as the ratio of per-nucleon DIS structure functions in nuclei $A$
and deuteron $D$, $R^A = F_2^A(x,Q^2)/F_2^D(x,Q^2)$, or equivalently, the ratio of
corresponding parton densities. This ratio exhibits distinct behaviors across
different kinematic regions: shadowing ($x \leq 0.1$), anti-shadowing ($0.1 \leq x
\leq 0.3$), valence quark dominance ($0.3 \leq x \leq 0.7$), and Fermi motion ($x
\geq 0.7$). Shadowing and anti-shadowing correspond to $R^A < 1$ and $R^A > 1$,
respectively, while the EMC effect \cite{EMCEffect} and Fermi motion refer to the
slope of $R^A$ in the valence-dominant region and the rise of $R^A$ at large $x$.

Note that the nuclear medium effect was first discovered \cite{EMCEffect} by the
European Muon Collaboration (EMC) in the valence quark dominance region.

Within the framework of the $Q^2$-rescaling model (hereafter referred to as the
rescaling model), nuclear medium modifications have been investigated
\cite{RescalingModel-1,RescalingModel-2,RescalingModel-Apps1,RescalingModel-Apps2}.
Initially
proposed for the valence-dominant region, this model suggests that the effective
confinement size of gluons and quarks in the nucleus is larger than in the free
nucleon. This shift in the confinement scale leads to a relation between PDFs and
nPDFs through a simple rescaling of their arguments \cite{RescalingModel-1,
RescalingModel-2}. Recently, the rescaling model has been extended to the low-$x$
region \cite{RescalingModel-Apps1}. It has been shown \cite{RescalingModel-Apps1,
AbKoLi23, RescalingModel-Apps2} that good agreement with available experimental data
on $F_2^A(x,Q^2)/F_2^D(x,Q^2)$ ratios at low and moderate $x$ can be achieved by
fitting the corresponding rescaling parameters.

The aim of this work is to present the results of our recent papers
\cite{KL2025,KL2025res}. First, in \cite{KL2025}, we derived analytical expressions
for PDFs from the solution of the DGLAP evolution equations at leading order (LO) in
the QCD coupling. These expressions are based on exact asymptotics at small and
large $x$ (see, e.g., \cite{gDAS1,gDAS2} and \cite{PDFs-early-approach}, respectively) and
include subasymptotic terms constrained by momentum conservation and/or the
Gross-Llewellyn-Smith and Gottfried sum rules. Several phenomenological parameters
were determined from a rigorous fit to precision BCDMS, H1, and ZEUS experimental
data on the proton structure function $F_2(x,Q^2)$ over a wide kinematic region: $2
\cdot 10^{-5} \leq x \leq 0.75$ and $1.2 \leq Q^2 \leq 30000$ GeV$^2$. In
\cite{KL2025res}, we employed the solution for PDFs from \cite{KL2025} and used the
rescaling model \cite{RescalingModel-1,RescalingModel-2} to study their nuclear modifications,
additionally incorporating Fermi motion effects. To extract the rescaling
parameters, we performed a global fit to available experimental data on the ratios
$F_2^A(x,Q^2)/F_2^{D}(x,Q^2)$ for different nuclear targets $A$ collected by the
EMC, NMC, SLAC, BCDMS, JLab, and CLAS Collaborations (see \cite{KL2025res} and
references therein).

\section{Structure function $F_2(x,Q^2)$ and PDFs in a proton} \indent

We begin with some basic formulas used in our calculations. It is well known that
the proton structure function $F_2(x,Q^2)$ at LO in the QCD coupling can be
expressed as:
\begin{gather}
  F_2(x,Q^2) = \sum_{i=1}^{N_f} e_i^2 \left[f_{q_i}(x,Q^2) + f_{\bar{q}_i}(x,Q^2)
\right],
\label{S1.14}
\end{gather}
\noindent
where $e_i$ is the fractional electric charge of quark $q_i$, $N_f$ is the number of
active quark flavors, and $f_{q_i}(x,Q^2)$ and $f_{\bar q_i}(x,Q^2)$ represent the
quark and antiquark densities in a proton (multiplied by $x$), respectively. In the
fixed-flavor-number scheme (FFNS) with $N_f = 4$, where $b$ and $t$ quarks are
separated out, we have:
\begin{gather}
  F_2(x,Q^2) = \frac{5}{18} \, f_{SI}(x,Q^2) + \frac{1}{6} f_{NS}(x,Q^2),
\label{S1.15}
\end{gather}
\noindent
where the singlet part $f_{SI}(x,Q^2)$ includes the valence and sea quark components:
\begin{gather}
        f_{V}(x,Q^2) = f_u^V(x,Q^2) + f_d^V(x,Q^2), \quad f_{S}(x,Q^2) = \sum_{i=1}^{4}
\left[f_{q_i}^S(x,Q^2) + f_{\bar{q}_i}^S(x,Q^2) \right], \nonumber \\
        f_{SI}(x,Q^2) = \sum_{i=1}^{4} \left[f_{q_i}(x,Q^2) + f_{\bar{q}_i}(x,Q^2) \right]
= f_{V}(x,Q^2) + f_{S}(x,Q^2).
        \label{eq-FSI}
\end{gather}
\noindent
The nonsinglet part $f_{NS}(x,Q^2)$ contains the difference between up and down quarks:
\begin{gather}
    \hspace{-0.5cm}    f_{NS}(x,Q^2) = \sum_{q=u,\,c} \left[f_{q}(x,Q^2) + f_{\bar{q}}(x,Q^2) \right] -
\sum_{q=d,\,s} \left[f_{q}(x,Q^2) + f_{\bar{q}}(x,Q^2) \right].
        \label{eq-FNS-NF4}
\end{gather}
\noindent

These formulas are useful for the accurate determination of phenomenological
parameters in the PDF parameterizations derived in our papers
\cite{KL2025, PDFs-our}. The approach \cite{KL2025, PDFs-our} follows the idea
presented in \cite{PDFs-early-approach,PDFs-our-previous} and consists of two main
steps. First, we analytically compute the asymptotics of the solutions to the DGLAP
equations for parton densities at low and high values of the Bjorken variable $x$.
Second, to obtain analytical expressions for PDFs over the entire $x$ range, we
combine these solutions and interpolate between them.

\section{Model of nuclear modifications} \indent

As a model for nuclear modifications of parton densities, we consider a combination
of the rescaling model \cite{RescalingModel-1,RescalingModel-2} and Fermi motion. The rescaling model
establishes a simple relationship between ordinary PDFs and nPDFs through a shift in
the kinematic variable $Q^2$.
For a nucleus $A$,
the valence and nonsinglet parts are modified as:
\begin{gather}
  f_{i}^A(x,Q^2) = f_{i}(x,Q^2_{A,\,i}),
  \label{va.1a}
\end{gather}
\noindent
where $i = V$ or $NS$, and the scale $Q^2_{A,\,i}$ is related to $Q^2$ by:
\begin{gather}
        s^A_i \equiv \ln \left(\frac{\ln\left(Q^2_{A,\,i}/\Lambda^2_{\rm
QCD}\right)}{\ln\left(Q^2_{0}/\Lambda^2_{\rm QCD}\right)}\right) = s +\ln
\left(1+\delta^A_i \right),
        \label{sA}
\end{gather}
\noindent
where $\delta^A_i$ are scale-independent free parameters.

\section{Numerical results} \indent

In Ref. \cite{KL2025}, we began by determining the phenomenological parameters in
the derived analytical expressions for PDFs in a proton. We then studied their
nuclear modifications in \cite{KL2025res}.

{\bf 1.}~The analytical expressions for PDFs obtained in \cite{KL2025} contain
several parameters that are not predicted by theory. Some of these parameters,
essential in the large-$x$ region, were determined by direct comparison with
numerical solutions of the DGLAP equations at the starting scale $Q_0^2$. We used
the CT'14 (LO) parameterizations \cite{CT14} from the CTEQ-TEA group with $N_f = 4$,
which matches our setup. The remaining parameters were determined by fitting precise
data on the proton structure function $F_2(x,Q^2)$ from the BCDMS, H1, and ZEUS
Collaborations (see \cite{KL2025} and references therein). These data cover an
extremely wide range of $x$ and $Q^2$: $2 \cdot 10^{-5} < x < 0.75$ and $1.2 < Q^2 <
30000$ GeV$^2$, allowing us to extract all necessary parameters simultaneously. Our
fit, comprising a total of 933 points from 5 datasets, achieves an excellent
goodness-of-fit, with $\chi^2/d.o.f. = 1.408$.

\begin{figure}
\begin{center}
\includegraphics[width=6.7cm]{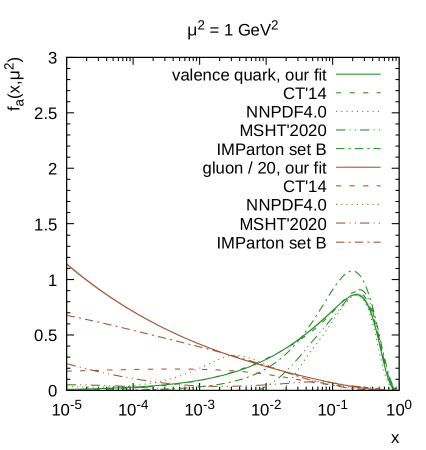}
\includegraphics[width=6.7cm]{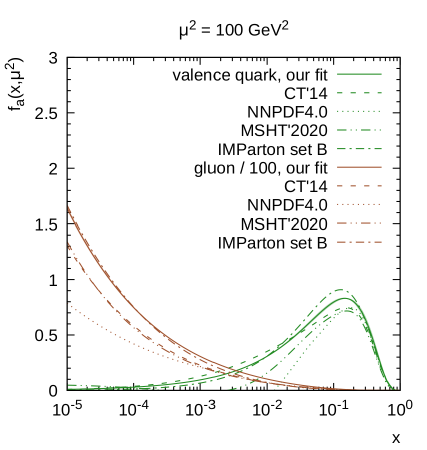}
\caption{Proton PDFs as functions of $x$ for different values of $Q^2$. For
comparison, we show the results of numerical solutions of the DGLAP equations from
the CTEQ-TEA \cite{CT14}, NNPDF4.0 \cite{NNPDF4}, MSHT'2020 \cite{MSHT20}, and IMP
\cite{IMP} groups.}
\label{fig2}
\end{center}
\end{figure}

Some of the derived PDFs are shown in Fig.~\ref{fig2} as functions of $x$ for
several scales $Q^2$, namely $Q^2 = 1$ and $100$ GeV$^2$. Additionally, we compare
our predictions with the corresponding FFNS results from the CTEQ-TEA Collaboration
\cite{CT14} at $N_f = 4$ and the variable-flavor-number scheme (VFNS) predictions
from the NNPDF4.0 \cite{NNPDF4}, MSHT'2020 \cite{MSHT20}, and IMP \cite{IMP} groups.
One can observe significant discrepancies between the different PDFs, especially at
low $Q^2 \sim 1$ GeV$^2$. Nevertheless, at moderate and higher $Q^2$, we find
reasonable agreement between our analytical results and the relevant numerical
analyses.

{\bf 2.} To extract the rescaling parameters
for various nuclear targets, in \cite{KL2025res} we performed a
global fit to structure function ratio data from the EMC, NMC, SLAC, BCDMS, E665,
JLab, and CLAS Collaborations (see \cite{KL2025res} and references therein). We
imposed kinematic cuts on the experimental data to ensure they lie in the deep
inelastic region: $Q^2 \geq 1$ GeV$^2$ and $W^2 \geq 4$ GeV$^2$. Following
\cite{KL2025}, we set $\Lambda_{\rm QCD}^{(4)} = 118$ MeV, corresponding to the
world average $\alpha_s(M^2_Z) = 0.1180$ \cite{PDG}. We applied a 'frozen' treatment
of the QCD coupling in the infrared region (see, for example, \cite{frozen-aQCD,
gDAS-PDFs-smallx} and references therein), where $\alpha_s(Q^2) \to \alpha_s(Q^2 +
M_\rho^2)$ with $M_\rho \sim 1$ GeV. This treatment leads to a good description of
the data on the proton structure function $F_2(x,Q^2)$ \cite{KL2025}.

\begin{figure}
        \begin{center}
                \includegraphics[width=5.7cm]{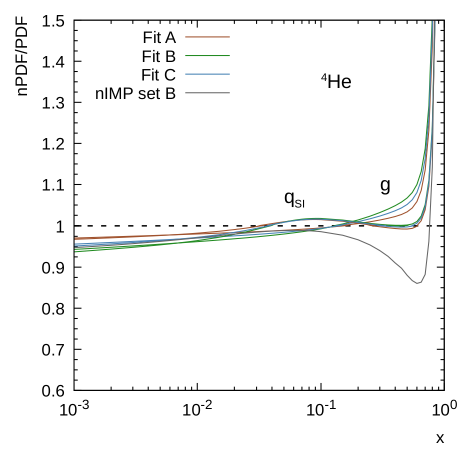}
                \includegraphics[width=5.7cm]{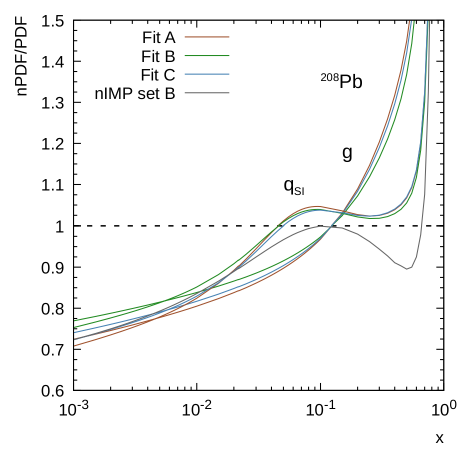}
                \caption{Predicted nuclear modification factors for parton distributions in
several nuclear targets. The results for gluon nuclear modification predicted by
the nIMP group \cite{nIMP} are shown for comparison.}
                \label{fig:3}
        \end{center}
\end{figure}

Using the analytical expressions for the nucleon target from \cite{KL2025},
Eqs.~(\ref{va.1a}) and~(\ref{sA}) for nuclear targets, and the fitted values of the
rescaling parameters (we considered three different forms of $A$-dependence), we can
predict nuclear modifications of PDFs. Our results are shown in Fig.~\ref{fig:3} for
several nuclei. As is well known, nuclear modifications are weakly dependent on
$Q^2$ at low and moderate $Q^2$, so we show only the results for $Q^2 = 10$ GeV$^2$.
For comparison, we also plot the results for gluon nuclear modification factors
predicted by the nIMP group \cite{nIMP}. We find that the shadowing effect for
gluons is generally weaker than for quarks, consistent with other studies (see, e.g.
\cite{nIMP, nCTEQ, nNNPDF3, EPPS21}). Nevertheless, the differences between
these results are rather small, and the predicted shadowing effects are close to the
nIMP expectations. This contrasts with the large-$x$ region, where we find a
significant discrepancy with the nIMP calculations. The latter show very weak
anti-shadowing at $x \sim 0.1$, while our calculations indicate significant
anti-shadowing effects for quarks. Note that other groups \cite{nNNPDF3, EPPS21}
predict very strong anti-shadowing for nuclear gluon densities. Thus, it is
currently difficult to draw specific conclusions. Again, precision measurements at
future colliders (EiC, EicC) are needed to clarify this point.

\section*{Conclusion} \indent

We have presented a brief overview of the results obtained in our recent papers
\cite{KL2025,KL2025res}. In \cite{KL2025}, we derived an analytical solution of the
DGLAP equations for PDFs in a proton, where the valence and non-singlet quark
components satisfy the Gross-Llewellyn-Smith and Gottfried sum rules, and momentum
conservation for the singlet quark and gluon densities is maintained. In
\cite{KL2025res}, we studied nuclear medium modifications of parton densities based
on the rescaling model. Furthermore, we extended the rescaling model to include
Fermi motion effects, providing a consistent description of nuclear modifications
for DIS structure functions and PDFs across the entire kinematic range. Using simple
analytical formulas for proton PDFs derived at LO in the QCD coupling, we performed
a global analysis of available deep inelastic data for different nuclear targets and
extracted the corresponding rescaling parameters. Our results highlight distinct
shadowing and anti-shadowing behaviors for gluons and quarks.

\section*{Acknowledgements} \indent

We thank S.P.~Baranov, H.~Jung, M.A. Malyshev, and N.N. Nikolaev for their interest,
valuable comments, and remarks. This research was supported by the Russian Science
Foundation grant No. 25-22-00066, https://rscf.ru/en/project/25-22-00066/.

\bibliography{EMCiN}
\end{document}